\documentclass[aps,prd,superscriptaddress,showpacs,floatfix,twocolumn]{revtex4}

\usepackage{graphicx}

\begin{document}

\title{Improved Measurement of Double Helicity Asymmetry 
in Inclusive Midrapidity $\pi^{0}$ Production 
for Polarized $p+p$ Collisions at $\sqrt{s}=200$ GeV}

\newcommand{\abilene}{Abilene Christian University, Abilene, TX 79699, U.S.}
\newcommand{\acadsin}{Institute of Physics, Academia Sinica, Taipei 11529, Taiwan}
\newcommand{\banaras}{Department of Physics, Banaras Hindu University, Varanasi 221005, India}
\newcommand{\barc}{Bhabha Atomic Research Centre, Bombay 400 085, India}
\newcommand{\bnl}{Brookhaven National Laboratory, Upton, NY 11973-5000, U.S.}
\newcommand{\caucr}{University of California - Riverside, Riverside, CA 92521, U.S.}
\newcommand{\ciae}{China Institute of Atomic Energy (CIAE), Beijing, People's Republic of China}
\newcommand{\cns}{Center for Nuclear Study, Graduate School of Science, University of Tokyo, 7-3-1 Hongo, Bunkyo, Tokyo 113-0033, Japan}
\newcommand{\colorado}{University of Colorado, Boulder, CO 80309, U.S.}
\newcommand{\columbia}{Columbia University, New York, NY 10027 and Nevis Laboratories, Irvington, NY 10533, U.S.}
\newcommand{\dapnia}{Dapnia, CEA Saclay, F-91191, Gif-sur-Yvette, France}
\newcommand{\debrecen}{Debrecen University, H-4010 Debrecen, Egyetem t{\'e}r 1, Hungary}
\newcommand{\elte}{ELTE, E{\"o}tv{\"o}s Lor{\'a}nd University, H - 1117 Budapest, P{\'a}zm{\'a}ny P. s. 1/A, Hungary}
\newcommand{\fsu}{Florida State University, Tallahassee, FL 32306, U.S.}
\newcommand{\gsu}{Georgia State University, Atlanta, GA 30303, U.S.}
\newcommand{\hiroshima}{Hiroshima University, Kagamiyama, Higashi-Hiroshima 739-8526, Japan}
\newcommand{\ihepprot}{IHEP Protvino, State Research Center of Russian Federation, Institute for High Energy Physics, Protvino, 142281, Russia}
\newcommand{\illuiuc}{University of Illinois at Urbana-Champaign, Urbana, IL 61801, U.S.}
\newcommand{\isu}{Iowa State University, Ames, IA 50011, U.S.}
\newcommand{\jinrdubna}{Joint Institute for Nuclear Research, 141980 Dubna, Moscow Region, Russia}
\newcommand{\kaeri}{KAERI, Cyclotron Application Laboratory, Seoul, South Korea}
\newcommand{\kek}{KEK, High Energy Accelerator Research Organization, Tsukuba, Ibaraki 305-0801, Japan}
\newcommand{\kfki}{KFKI Research Institute for Particle and Nuclear Physics of the Hungarian Academy of Sciences (MTA KFKI RMKI), H-1525 Budapest 114, POBox 49, Budapest, Hungary}
\newcommand{\korea}{Korea University, Seoul, 136-701, Korea}
\newcommand{\kurchatov}{Russian Research Center ``Kurchatov Institute", Moscow, Russia}
\newcommand{\kyoto}{Kyoto University, Kyoto 606-8502, Japan}
\newcommand{\labllr}{Laboratoire Leprince-Ringuet, Ecole Polytechnique, CNRS-IN2P3, Route de Saclay, F-91128, Palaiseau, France}
\newcommand{\lawllnl}{Lawrence Livermore National Laboratory, Livermore, CA 94550, U.S.}
\newcommand{\losalamos}{Los Alamos National Laboratory, Los Alamos, NM 87545, U.S.}
\newcommand{\lpc}{LPC, Universit{\'e} Blaise Pascal, CNRS-IN2P3, Clermont-Fd, 63177 Aubiere Cedex, France}
\newcommand{\lund}{Department of Physics, Lund University, Box 118, SE-221 00 Lund, Sweden}
\newcommand{\muenster}{Institut f\"ur Kernphysik, University of Muenster, D-48149 Muenster, Germany}
\newcommand{\myongji}{Myongji University, Yongin, Kyonggido 449-728, Korea}
\newcommand{\nagasaki}{Nagasaki Institute of Applied Science, Nagasaki-shi, Nagasaki 851-0193, Japan}
\newcommand{\newmex}{University of New Mexico, Albuquerque, NM 87131, U.S. }
\newcommand{\nmsu}{New Mexico State University, Las Cruces, NM 88003, U.S.}
\newcommand{\ornl}{Oak Ridge National Laboratory, Oak Ridge, TN 37831, U.S.}
\newcommand{\orsay}{IPN-Orsay, Universite Paris Sud, CNRS-IN2P3, BP1, F-91406, Orsay, France}
\newcommand{\peking}{Peking University, Beijing, People's Republic of China}
\newcommand{\pnpi}{PNPI, Petersburg Nuclear Physics Institute, Gatchina, Leningrad region, 188300, Russia}
\newcommand{\riken}{RIKEN, The Institute of Physical and Chemical Research, Wako, Saitama 351-0198, Japan}
\newcommand{\rikjrbrc}{RIKEN BNL Research Center, Brookhaven National Laboratory, Upton, NY 11973-5000, U.S.}
\newcommand{\rikkyo}{Physics Department, Rikkyo University, 3-34-1 Nishi-Ikebukuro, Toshima, Tokyo 171-8501, Japan}
\newcommand{\saispbstu}{Saint Petersburg State Polytechnic University, St. Petersburg, Russia}
\newcommand{\saopaulo}{Universidade de S{\~a}o Paulo, Instituto de F\'{\i}sica, Caixa Postal 66318, S{\~a}o Paulo CEP05315-970, Brazil}
\newcommand{\seoulnat}{System Electronics Laboratory, Seoul National University, Seoul, South Korea}
\newcommand{\stonybrkc}{Chemistry Department, Stony Brook University, Stony Brook, SUNY, NY 11794-3400, U.S.}
\newcommand{\stonycrkp}{Department of Physics and Astronomy, Stony Brook University, SUNY, Stony Brook, NY 11794, U.S.}
\newcommand{\subatech}{SUBATECH (Ecole des Mines de Nantes, CNRS-IN2P3, Universit{\'e} de Nantes) BP 20722 - 44307, Nantes, France}
\newcommand{\tenn}{University of Tennessee, Knoxville, TN 37996, U.S.}
\newcommand{\titech}{Department of Physics, Tokyo Institute of Technology, Oh-okayama, Meguro, Tokyo 152-8551, Japan}
\newcommand{\tsukuba}{Institute of Physics, University of Tsukuba, Tsukuba, Ibaraki 305, Japan}
\newcommand{\vandy}{Vanderbilt University, Nashville, TN 37235, U.S.}
\newcommand{\waseda}{Waseda University, Advanced Research Institute for Science and Engineering, 17 Kikui-cho, Shinjuku-ku, Tokyo 162-0044, Japan}
\newcommand{\weizmann}{Weizmann Institute, Rehovot 76100, Israel}
\newcommand{\yonsei}{Yonsei University, IPAP, Seoul 120-749, Korea}
\affiliation{\abilene}
\affiliation{\acadsin}
\affiliation{\banaras}
\affiliation{\barc}
\affiliation{\bnl}
\affiliation{\caucr}
\affiliation{\ciae}
\affiliation{\cns}
\affiliation{\colorado}
\affiliation{\columbia}
\affiliation{\dapnia}
\affiliation{\debrecen}
\affiliation{\elte}
\affiliation{\fsu}
\affiliation{\gsu}
\affiliation{\hiroshima}
\affiliation{\ihepprot}
\affiliation{\illuiuc}
\affiliation{\isu}
\affiliation{\jinrdubna}
\affiliation{\kaeri}
\affiliation{\kek}
\affiliation{\kfki}
\affiliation{\korea}
\affiliation{\kurchatov}
\affiliation{\kyoto}
\affiliation{\labllr}
\affiliation{\lawllnl}
\affiliation{\losalamos}
\affiliation{\lpc}
\affiliation{\lund}
\affiliation{\muenster}
\affiliation{\myongji}
\affiliation{\nagasaki}
\affiliation{\newmex}
\affiliation{\nmsu}
\affiliation{\ornl}
\affiliation{\orsay}
\affiliation{\peking}
\affiliation{\pnpi}
\affiliation{\riken}
\affiliation{\rikjrbrc}
\affiliation{\rikkyo}
\affiliation{\saispbstu}
\affiliation{\saopaulo}
\affiliation{\seoulnat}
\affiliation{\stonybrkc}
\affiliation{\stonycrkp}
\affiliation{\subatech}
\affiliation{\tenn}
\affiliation{\titech}
\affiliation{\tsukuba}
\affiliation{\vandy}
\affiliation{\waseda}
\affiliation{\weizmann}
\affiliation{\yonsei}
\author{S.S.~Adler}	\affiliation{\bnl}
\author{S.~Afanasiev}	\affiliation{\jinrdubna}
\author{C.~Aidala}	\affiliation{\columbia}
\author{N.N.~Ajitanand}	\affiliation{\stonybrkc}
\author{Y.~Akiba}	\affiliation{\kek}  \affiliation{\riken}  \affiliation{\rikjrbrc}
\author{A.~Al-Jamel}	\affiliation{\nmsu}
\author{J.~Alexander}	\affiliation{\stonybrkc}
\author{K.~Aoki}	\affiliation{\kyoto} \affiliation{\riken}
\author{L.~Aphecetche}	\affiliation{\subatech}
\author{R.~Armendariz}	\affiliation{\nmsu}
% ...
\collaboration{PHENIX Collaboration} \noaffiliation

\date{\today}

%

\begin{abstract}
We present an improved measurement of the double helicity asymmetry
for $\pi^{0}$ production in polarized proton-proton scattering at
$\sqrt{s}=200$~GeV employing the PHENIX detector at the
Relativistic Heavy Ion Collider (RHIC).  The improvements to our
previous measurement come from two main factors:  Inclusion of a
new data set from the 2004 RHIC run with higher beam polarizations
than the earlier run and a recalibration of the beam polarization 
measurements, which resulted in reduced uncertainties and increased beam
polarizations.  The results are compared to a Next to Leading Order
(NLO) perturbative Quantum Chromodynamics (pQCD) calculation with a
range of polarized gluon distributions.
\end{abstract}
\maketitle


From polarized lepton-nucleon deep inelastic scattering (DIS) 
experiments it is known that only $\sim 25\%$ of the proton
spin can be attributed to the spins of the quarks and 
anti-quarks~\cite{DISsigma}.  The rest of the proton spin 
is assumed to be carried by the gluons and orbital angular momentum. 

The double helicity asymmetry of inclusive $\pi^{0}$ production
in polarized $p+p$ collisions, $A_{LL}^{\pi^{0}}$, is directly 
sensitive to the polarized gluon distribution function in the proton
through gluon-gluon and gluon-quark subprocesses~\cite{theory}.
Results on $A_{LL}^{\pi^{0}}$ from polarized $p+p$ collisions
at $\sqrt{s}=$200 GeV using the PHENIX detector at the Relativistic
Heavy Ion Collider (RHIC) were published earlier~\cite{pi0r3}.
The unpolarized cross section for $\pi^{0}$
production is described well by next-to-leading-order
perturbative QCD (NLO pQCD) calculations within the theoretical scale
uncertainty~\cite{pi0r2}.    The double spin asymmetry can
hence be reliably interpreted in terms of the polarized gluon and 
polarized quark distributions in the nucleon.

$A_{LL}^{\pi^{0}}$ is the difference between $\pi^{0}$ production 
cross sections for 
like helicity and unlike helicity proton collisions, divided by the sum. 
Experimentally, $A_{LL}^{\pi^{0}}$ is obtained from the 
difference in production rates for polarized collisions of like and 
unlike helicities, divided by the sum, and normalized by the product 
of the polarizations of each beam.

The published $A_{LL}^{\pi^{0}}$ data~\cite{pi0r3} were collected 
in the 2003 RHIC run (Run-3) when the average polarization 
of the beams in RHIC was 0.35 and 0.30 
for the ``blue'' and ``yellow'' ring, respectively. 
Recalibration of the proton-carbon CNI polarimeter~\cite{pcPol}, based on a 
polarized atomic hydrogen gas jet at RHIC~\cite{jetPol}, 
changed the absolute scale of beam polarizations by 22\%.
In addition, this recalibration reduced the uncertainty in the polarization 
measurement for each beam from $\sim34\%$ to $\sim18\%$ 
(and for a product of two beam polarizations from 65\% to 28\%).
An important, but less significant, change in Run-3 data as 
published occurred  
when we found a polarization pattern error in 10\% of the data 
files of Run-3, and this was fixed during this reanalysis.

In 2004, one week of RHIC operation was dedicated to the study and
improvement of the beam polarization in RHIC. The PHENIX detector took
data during this time. The results of this run (Run-4) are presented here
combined with the data from the previous year.


The statistical uncertainty of a double spin 
asymmetry is proportional to the square root of the figure of merit defined 
by  ${\cal L}P^{2}_{B}P^{2}_{Y}$,  where the $P_{B}$ and $P_{Y}$ are the beam 
polarization values of the blue and yellow RHIC beams, respectively 
and ${\cal L}$ is the integrated luminosity collected by the experiment. 
Table \ref{tab:compare} shows the average beam polarization values, the 
integrated luminosities and the figures of merit of the two runs.

\begin{table}[hbt]
\caption{\label{tab:compare} Comparison of the data sets from 2003 (Run-3)
and 2004 (Run-4) RHIC runs. }
\begin{ruledtabular}
\begin{tabular}{lcccc}
Run& $\langle P_{B} \rangle$ & $\langle P_{Y} \rangle$ & ${\cal L}$ (nb$^{-1}$) & ${\cal L} \langle P_{B} P_{Y} \rangle ^{2} $ (nb$^{-1}$) \\
\hline
Run-3& 0.35 & 0.30 & 220 & 2.6 \\
Run-4& 0.45 & 0.44 &  75 & 2.9
\end{tabular}
\end{ruledtabular}
\end{table}

\begin{table}[hbt]
\caption{\label{tab:tab2} $A_{LL}^{\pi^{0}}$ for four $p_{T}$ bins. 
The total point-to-point uncertainties are shown. 
Scale uncertainties of 28\%, 24\% and 18\% for Run-3, Run-4 and 
combined runs are not included.  For the combined results, 
uncorrelated systematic uncertainties for Run-3 and Run-4 
have been included as point-to-point uncertainties, 
and are shown in parenthesis. }
\begin{ruledtabular}
\begin{tabular}{cccc}
p$_{T}$ ($\langle$ p$_{T}$ $\rangle$) & Run-3 & Run-4 & Run-3+Run-4 \\
(GeV/c) & ($10^{-2}$) & ($10^{-2}$) & ($10^{-2}$) \\
\hline
1-2 (1.59) & -0.97$\pm$0.85 & -0.35$\pm$0.72 & -0.60$\pm$0.56 (0.10) \\
2-3 (2.39) & -0.92$\pm$0.90 &  0.56$\pm$0.82 & -0.10$\pm$0.61 (0.10) \\
3-4 (3.37) & -0.90$\pm$1.88 &  0.03$\pm$1.79 & -0.41$\pm$1.30 (0.09) \\
4-5 (4.38) & -1.37$\pm$4.19 &  7.90$\pm$4.04 &  3.23$\pm$2.98 (0.65) \\
\end{tabular}
\end{ruledtabular}
\end{table}

The stable spin direction in RHIC is vertical, and spin rotators 
are used to provide longitudinal polarization at PHENIX~\cite{rot}. 
For the Run-3 and Run-4 data sets, the longitudinal fraction of the 
beam polarization was $>0.98$  and $>0.99$, respectively.
Details of the Run-4 analysis method including the handling of 
detector efficiency, relative luminosity, determination of the 
longitudinal fraction of the beam polarization in the PHENIX 
interaction region, and the background subtraction are 
described in our Run-3 publication~\cite{pi0r3}. 

Table~\ref{tab:tab2} shows the results of the Run-3 reanalysis and the 
Run-4 analysis.  Bunch-to-bunch and fill-to-fill systematic uncertainties 
for the asymmetry measurements are negligible. Total scale uncertainty due 
to the uncertainty in beam polarization is given in the Table caption for each 
data set and combined data. The beam polarization uncertainty in each data set 
consists of two terms: an 18\% scale uncertainty (9\% for each beam), 
due to the proton-carbon CNI polarimeter calibration, common for each data 
set; and an uncorrelated uncertainty (21\% in Run-3 and 16\% in Run-4), 
that was included in the point-to-point uncertainties for the combined result.
The additional $p_T$-independent systematic uncertainty related to the 
relative luminosity measurement between colliding bunches with the 
same and opposite helicities was estimated to be below $1.5\times10^{-3}$.

\begin{figure}[hbt]
\includegraphics[width=1.0\linewidth]{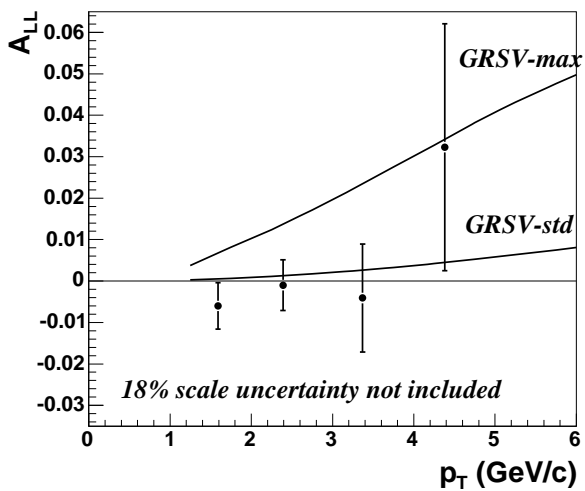}
\caption{\label{fig:all} Run-3+Run-4 combined results on 
$A_{LL}^{\pi^{0}}$ versus mean $p_{T}$ in each bin. 
Two theoretical calculations based on NLO pQCD are also shown for 
comparison with the data (see text and~\protect\cite{theory,theory2} for 
details).}
\end{figure}

Table~\ref{tab:tab2} and Fig.~\ref{fig:all} show the combined asymmetries
for Run-3 and Run-4 data and that the two results are consistent within
their statistical uncertainties. Fig.~\ref{fig:all} also presents two NLO
pQCD calculations using different assumptions for the polarized
gluon distribution.  One uses the best global fit to the inclusive DIS
data (GRSV-std) and the other uses a polarized gluon distribution equal to
the unpolarized distribution at the input scale of $Q^{2}=0.4$ GeV$^{2}$
(GRSV-max)~\cite{theory,theory2}.

Following the discussion in~\cite{pi0r3}, we compared our results with 
theoretical curves, calculating confidence levels (C.L.) for all $p_T$ 
points ($>$1 GeV/$c$) and for the three highest $p_T$ points ($>$2 GeV/$c$). 
The data are consistent with GRSV-std with a C.L. = 61-63\% (73-74\%) 
for the range in polarization uncertainty of the measurement, 
considering all four points (the three highest $p_T$ points).
The results are less consistent with the large gluon polarization, 
with C.L. = 0.04-2\% (0.3-5\%) for GRSV-max~\cite{theory,theory2}.
These confidence levels do not include the theoretical uncertainty 
from the choice of scales or of the parton distribution functions 
or fragmentation function.  

In summary, we present results for $A_{LL}^{\pi^{0}}$ that are
significantly improved over our previous publication~\cite{pi0r3}.
Both the statistical and systematic uncertainties are reduced by
more than a factor of two.  The new results probe the polarized gluon
distribution in the proton with considerably improved resolving power. 
We conclude that the observed asymmetry is small and not consistent 
with a maximal gluon polarization.


We thank the staff of the Collider-Accelerator and 
Physics Departments at BNL for their vital contributions.  
We acknowledge support from 
the Department of Energy and NSF (U.S.A.), 
MEXT and JSPS (Japan), 
CNPq and FAPESP (Brazil), 
NSFC (China), 
IN2P3/CNRS, and CEA (France), 
BMBF, DAAD, and AvH (Germany), 
OTKA (Hungary), 
DAE (India), 
ISF (Israel), 
KRF, CHEP, and KOSEF (Korea), 
MES, RAS, and FAAE (Russia),
VR and KAW (Sweden), 
U.S. CRDF for the FSU, 
US-Hungarian NSF-OTKA-MTA, 
and US-Israel BSF.


\begin{thebibliography}{30}
\bibitem{DISsigma} J. Ashman {\em et al.}, Phys. Lett. {\bf B206}, 
                   364 (1988), Nucl. Phys. {\bf B328}, 1 (1989);
              E. Hughes and R. Voss, Ann. Rev. Nucl. Part. Sci. {\bf 49}, 
              303 (1999). 

\bibitem{theory} B.~J\"ager, A.~Sch\"{a}fer, M.~Stratmann, and 
W.~Vogelsang, \PRD {\bf 67}, 054005 (2003).

\bibitem{pi0r3} S.S.~Adler {\it et al.}, \PRL {\bf 93}, 202002 (2004).

\bibitem{pi0r2} S.S.~Adler {\it et al.}, \PRL {\bf 91}, 241803 (2003).

\bibitem{pcPol} O.~Jinnouchi {\it et al.}, 
RHIC/CAD Accelerator Physics Note 171 (2004).

\bibitem{jetPol} T.~Wise {\it et al.} and H.~Okada {\it et al.}, 
Proceedings of the Spin 2004 Symposium, p.757 and p.507. 

\bibitem{rot} W.W.~MacKay {\it et al.}, Proceedings of 
the 2003 Particle Accelerator Conference, p.1697.

\bibitem{theory2} M.~Gl\"uck, E.~Reya, M.~Stratmann, and 
W.~Vogelsang, \PRD {\bf 63}, 094005 (2001).

\end{thebibliography}

\def\PRL{Phys. Rev. Lett.\ }
\def\PRD{{Phys. Rev.}~{\bf D}}

\end{document}